\documentclass[fleqn,twoside]{article}
\usepackage{espcrc2,epsf}

\usepackage{graphicx}
\usepackage[figuresright]{rotating}

\newcommand{\AmS}{{\protect\the\textfont2
  A\kern-.1667em\lower.5ex\hbox{M}\kern-.125emS}}

\title{Factorizing one-loop contributions to two-loop Bhabha scattering
        and automatization of Feynman diagram calculations}

\author{J.~Fleischer\address{Universit\"at Bielefeld, 
        Fakult\"at f\"ur Physik, 
        Universit\"atsstr. 25, 33615 Bielefeld, Germany}
        \thanks{J.~Fleischer likes to thank 
        CERN-INTAS for financial support under project no. INTAS-CERN 99-0377},%
        O.~V.~ Tarasov\addressmark\thanks{O.V. Tarasov likes to thank the
        DFG for financial support under project no. FL 241/4-2},
        T.~Riemann\address[DESY]{DESY Zeuthen, Platanenallee 6, 15738 Zeuthen, Germany}
        and
        A.~Werthenbach\addressmark[DESY]}
       
\begin{document}

\begin{abstract}
   In higher order calculations a number of new technical problems arise:
one needs diagrams in arbitrary dimension in order to obtain their needed
$\varepsilon$-expansion, zero Gram determinants appear,
renormalization produces diagrams with `dots' on the lines, i.e. higher
order powers of scalar propagators. All these problems cannot be accessed
by the `standard' Passarino-Veltman approach: there is not available
what is needed for higher loops. We demonstrate our method of how to solve
these problems.
\vspace{1pc}
\end{abstract}

\maketitle

\section{DIANA}

  We are moving in the direction of {\bf two-loop} Bhabha scattering, which
is extremely important, in particular at higher energies, for the luminosity
determination of the coming accellerators. By means of {\bf DIANA} 
\cite{Tentyukov:1999is,Tentyukov:1999yq,Fleischer:2000zr} we produced the 
following one-loop diagrams:

\begin{center}
\vbox{
 \raisebox{6.0cm}{\makebox[0pt]{\hspace*{-2cm}$$}}
 \epsfysize=75mm \epsfbox{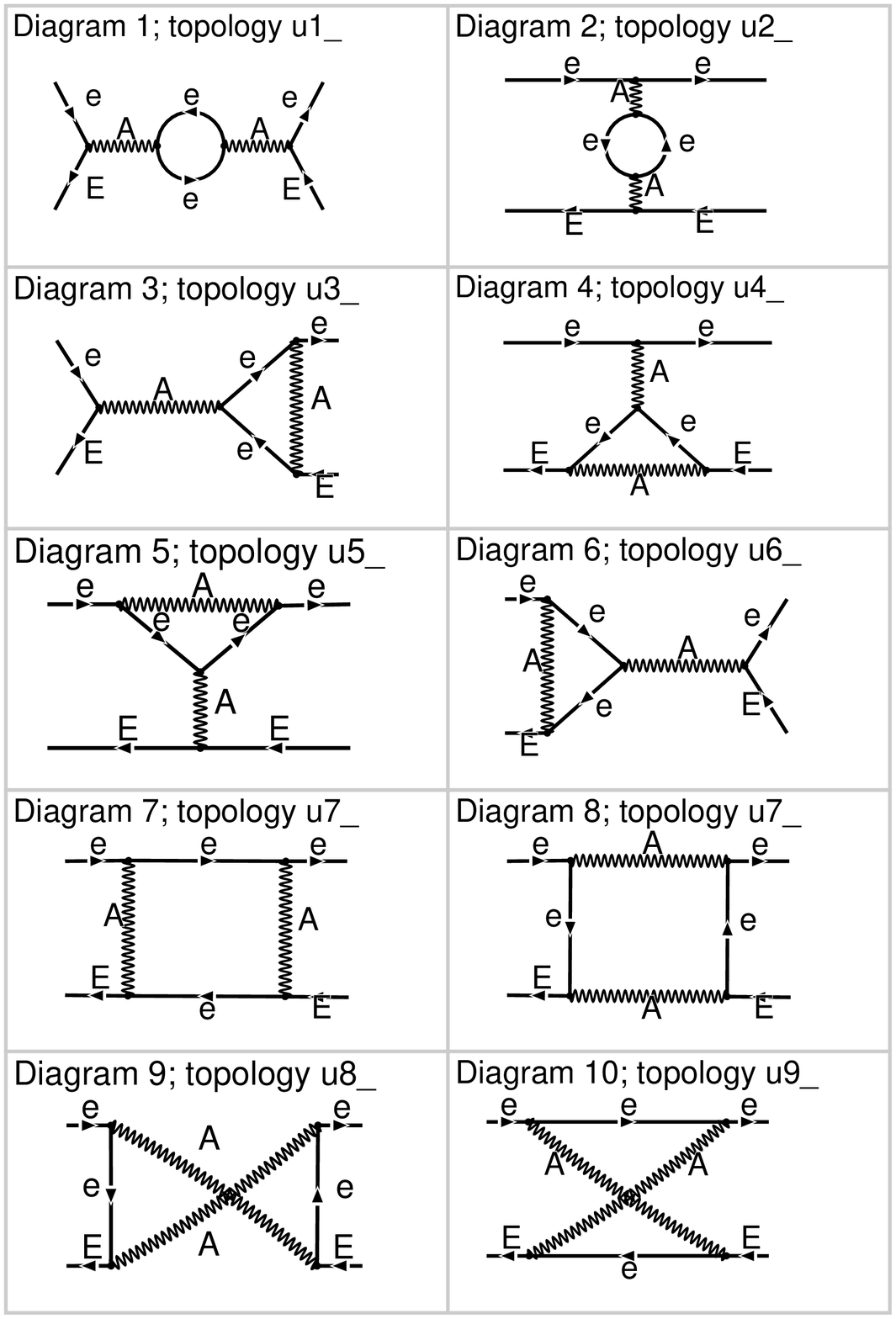}
     }
\end{center}

  Factorizing one-loop contributions are obtained e.g. in mass-renormalization: in
any Feynman diagram we replace
$$
\frac{1}{k^2-(m^2+\delta(m^2))} \to \frac{1}{k^2-m^2} (1+\frac{\delta(m^2)}{k^2-m^2}), \nonumber
$$
thus obtaining `dotted' diagrams.

\section{Calculation of 1-loop integrals}

The evaluation of one-loop integrals as used here was introduced 
in detail in \cite{Fleischer:1999hq}.
The method exists in first reducing tensor integrals to scalar ones in
higher dimension and higher indices (i.e. higher powers of scalar
propagators). Before further reducing these scalar integrals arready
the on-shell conditions can be implemented and the amplitude 
decomposition performed. The reduction to `master integrals' is performed
by the use of recursion relations. One of these, reducing the space 
time dimension, is given by:
\begin{eqnarray}
&&  (d-\sum_{i=1}^{n}\nu_i+1) \left(  \right)_n  I^{(d+2)}_n= \nonumber\\
&&  ~~~~~~~~~~~~~~~~~~~~
  \left[ {0 \choose 0}_n - \sum_{k=1}^n {0 \choose k}_n {\bf k^-} \right]I^{(d)}_n, \nonumber
\end{eqnarray}
where the objects ${i \choose j}_n$ are Gram determinants. As a particular problem
occurred in this connection zero Gram determinants , 
i.e. ${0 \choose 0}_n=0:$
\newpage
$$
  (d-\sum_{i=1}^{n}\nu_i+1) \left(  \right)_n  I^{(d+2)}_n=
   - \sum_{k=1}^n {0 \choose k}_n {\bf k^-} I^{(d)}_n.
$$
To increase again the dimension $d$, one then uses the relation
$$
\sum_{j=1}^n \nu_j {\bf j^+}I^{(d+2)}_n~ = - I^{(d)}_n~.
  \label{extra}
$$

\section{Decomposition of the diagrams into Amplitudes}

In order to avoid $\gamma_5$, we decompose the amplitdes in terms
of the following 12 structures:

\begin{displaymath}
\begin{array}{llclclcl}
O_1 =  I          \cdot   I   ,~
O_2 =  \hat{p}_4  \cdot   I   ,~
O_3 =  I          \cdot    \hat{p}_2   ,~\\
O_4 = \hat{p}_4   \cdot   \hat{p}_2    ,~
O_5 =  \gamma_{\mu} \cdot \gamma_{\mu}    ,~
O_6 =  \gamma_{\mu}\hat{p}_4  \cdot  \gamma_{\mu}   ,~\\
O_7 =  \gamma_{\mu}  \cdot   \gamma_{\mu}\hat{p}_2    ,~
O_8 =  \gamma_{\mu}\gamma_{\nu} \cdot   \gamma_{\nu}\gamma_{\mu}    ,~\\
O_9 =  \gamma_{\mu}\gamma_{\nu}\hat{p}_4  \cdot  \gamma_{\nu}\gamma_{\mu}    ,~
O_{10} =  \gamma_{\mu}\gamma_{\nu} \cdot  \gamma_{\nu}\gamma_{\mu}\hat{p}_2    ,~\\
O_{11} =  \gamma_{\mu}\gamma_{\nu}\hat{p}_4  \cdot \gamma_{\nu}\gamma_{\mu}\hat{p}_2  ,~
O_{12} =  \gamma_{\mu}\gamma_{\nu} \gamma_{\rho}  \cdot 
         \gamma_{\rho}\gamma_{\nu}\gamma_{\mu}    .
\end{array}
\end{displaymath}

Thus the on-shell diagrams read
$$
{\bf Diagram = \sum_{j=1}^{12} A_j O_j.} \nonumber
$$
with amplitudes ${\bf A_j}$. Finally we obtained `crossing relations' like
e.g. crossing diagrams 8 and 9 into each other:

\begin{eqnarray}
\lefteqn{A_1^{(9)} = A_1^{(8)}-2 m_e A_7^{(8)}-4 m_e A_9^{(8)}+2 d A_8^{(8)}} \nonumber \\
\lefteqn{A^{(9)}_2 = -A_2^{(8)}+4 A_7^{(8)}-2(d-4) A_9^{(8)}-8 m_e A_{11}^{(8)}}\nonumber \\
\lefteqn{A^{(9)}_3 = -A_3^{(8)}+2 A_6^{(8)}-2(d-2) A_{10}^{(8)}-4 m_e A_{11}^{(8)}}\nonumber \\
\lefteqn{A^{(9)}_4 = A_4^{(8)}-2(d-2) A_{11}^{(8)}}\nonumber \\
\lefteqn{A^{(9)}_5 = -A_5^{(8)}+2 m_e A_6^{(8)}+4 m_e A_7^{(8)}+8 m_e A_9^{(8)} }\nonumber \\
\lefteqn{~~~~~~~~~+4 m_e A_{10}^{(8)}-12 m_e^2 A_{11}^{(8)}-(6 d-4) A_{12}^{(8)}}\nonumber \\
\lefteqn{A^{(9)}_6 = -4 m_e A_{11}^{(8)}+A_6^{(8)}}\nonumber \\
\lefteqn{A^{(9)}_7 = -4 m_e A_{11}^{(8)}+A_7^{(8)}}\nonumber \\
\lefteqn{A^{(9)}_8 = -A_8^{(8)}}\nonumber \\
\lefteqn{A^{(9)}_9 = A_9^{(8)}}\nonumber \\
\lefteqn{A^{(9)}_{10} = A_{10}^{(8)}}\nonumber \\
\lefteqn{A^{(9)}_{11} = -A_{11}^{(8)}}\nonumber \\
\lefteqn{A^{(9)}_{12} = A_{12}^{(8)}}\nonumber
\end{eqnarray}
and {\bf exchanging  t and  u}. These relations also hold for 
crossing diagrams 7 to 10 (s and u exchanged). For the crossing of
7 and 8 (s and t exchanged) only a change of sign occurs in
the amplitudes. {\bf Thus: } we need to calculate {\bf only one} selfenergy, vertex and box.\\

{\bf Further relations} between amplitudes of {\bf one} diagram (obtained by
solving a system of equations) are:
\begin{eqnarray}
&&A_3 = A_2\nonumber \\
&&A_7 = A_6\nonumber \\
&&A_{10}= A_9\nonumber \\
&&A_8 = \frac{1}{4} A_1 - \frac{m_e}{2} A_2 +~(~A_{12}~) \nonumber \\
&&A_9 = \frac{1}{4 m_e} A_1 - \frac{1}{2} A_2 - \frac{1}{2} A_6 \nonumber \\
&&A_{11} = \frac{1}{4 m_e^2} A_1 - \frac{1}{2 m_e} A_2 \nonumber 
\end{eqnarray}
{\bf Thus:} there are only {\bf 6 indpendent amplitudes}.

The amplitudes for any of the diagrams, even for non-zero electron mass,
are relatively short. They were presented in the talk and can be obtained
from ACAT's webpage \cite{Fleischer:2002ac}.
What remains to do is to expand the master integrals, given
in arbitrary $d$, in $\varepsilon, (d=4-2 \varepsilon)$ in order to obtain properly
all finite contributions. This work is presently being done.

\end{document}